\documentclass[conference]{IEEEtran}
\IEEEoverridecommandlockouts
\usepackage[colorlinks=true,linkcolor=blue,urlcolor=blue,citecolor=blue,breaklinks=true]{hyperref}
\usepackage{amsmath,amssymb,amsfonts}
\renewcommand{\vec}[1]{\boldsymbol{#1}}
\usepackage{graphicx}
\usepackage[dvipsnames]{xcolor}

\makeatletter
\newcommand{\linebreakand}{%
  \end{@IEEEauthorhalign}
  \hfill\mbox{}\par
  \mbox{}\hfill\begin{@IEEEauthorhalign}
}
\makeatother

\def\BibTeX{{\rm B\kern-.05em{\sc i\kern-.025em b}\kern-.08em
    T\kern-.1667em\lower.7ex\hbox{E}\kern-.125emX}}

\begin{document}







\title{Non-linear, bivariate stochastic modelling of power-grid frequency applied to islands 
\thanks{We gratefully acknowledge funding from the Helmholtz Association under grant no. VH-NG-1727 and the Scientific Research Projects Coordination Unit of Istanbul University, Project no. 39071}}


\author{\IEEEauthorblockN{1\textsuperscript{st} Ulrich Oberhofer}
\IEEEauthorblockA{\textit{Institute for Automation and} \\
\textit{Applied Informatics}\\
\textit{Karlsruhe Institute of Technology}\\
Eggenstein-Leopoldshafen, Germany \\
ulrich.oberhofer@kit.edu
}
\and
\IEEEauthorblockN{2\textsuperscript{nd} Leonardo Rydin Gorj\~ao}
\IEEEauthorblockA{\textit{Faculty of Science and Technology}\\
\textit{Norwegian University of Life Sciences}\\
Ås, Norway \\
leo.rydin@nmbu.no}
\and
\IEEEauthorblockN{3\textsuperscript{rd} G. Cigdem Yalcin}
\IEEEauthorblockA{\textit{Department of Physics}\\
\textit{Istanbul University}\\
Istanbul, Turkey \\
gcyalcin@istanbul.edu.tr
\linebreakand
\IEEEauthorblockN{4\textsuperscript{th} Oliver Kamps}
\IEEEauthorblockA{\textit{Center for Nonlinear Sciences (CeNoS)} \\
\textit{University of Münster}\\
Münster, Germany\\
okamp@uni-muenster.de}
}
\and
\IEEEauthorblockN{5\textsuperscript{th} Veit Hagenmeyer}
\IEEEauthorblockA{\textit{Institute for Automation and} \\
\textit{Applied Informatics}\\
\textit{Karlsruhe Institute of Technology}\\
Eggenstein-Leopoldshafen, Germany \\
veit.hagenmeyer@kit.edu
}
\and
\IEEEauthorblockN{6\textsuperscript{th} Benjamin Schäfer}
\IEEEauthorblockA{\textit{Institute for Automation and} \\
\textit{Applied Informatics}\\
\textit{Karlsruhe Institute of Technology}\\
Eggenstein-Leopoldshafen, Germany \\
benjamin.schaefer@kit.edu
}
}


\maketitle

\begin{abstract}
Mitigating climate change requires a transition away from fossil fuels towards renewable energy. 
As a result, power generation becomes more volatile and options for microgrids and islanded power-grid operation are being broadly discussed.
Therefore, studying the power grids of physical islands, as a model for islanded microgrids, is of particular interest when it comes to enhancing our understanding of power-grid stability. 
In the present paper, we investigate the statistical properties of the power-grid frequency of three island systems: Iceland, Ireland, and the Balearic Islands.
We utilise a Fokker--Planck approach to 
construct stochastic differential equations that describe market activities, control, and noise acting on power-grid dynamics.
Using the obtained parameters we create synthetic time series of the frequency dynamics and compare them to empirical data. Our main contribution is to propose two extensions of stochastic power-grid frequency models and showcase the applicability of these new models to non-Gaussian statistics, as encountered in islands.
\end{abstract}


\maketitle
\makeatother

\section{Introduction} 

\subsection*{Motivation \& problem}
Controlling power-grid frequency is important in the design and operation of a stable power system. 
A shortage of power manifests itself in a decrease of the frequency and many control schemes to stabilise and balance the power system rely on frequency measurements~\cite{ Machowski2020}.
Hence, understanding power-grid frequency dynamics and statistics is critical. 

Obtaining such an understanding is not trivial: Power grids are complex systems driven by both stochastic and deterministic influences~\cite{Schaefer2018b, Schaefer2018c, RydinGorjao2020a}. 
Renewable generation~\cite{Anvari2016}, but also short-term consumer behaviour~\cite{Anvari2022, Han2022} are effectively random inputs to the power balance, while day-ahead trading, scheduled generation and overall demand trend are deterministic~\cite{Schaefer2018c}. 
Therefore, detailed model-based approaches that describe the dynamics of rotor-angle, angular velocity, and voltages~\cite{ Machowski2020} are complemented by data-driven approaches~\cite{RydinGorjao2020a, Kyesswa2020}.
Still, these approaches often focus on larger synchronous areas and not on small, island grids.

Understanding the statistics of power-grid-related variables in islands is particularly useful as case studies since geographical islands are often isolated and only (weakly) coupled via DC lines to other (continental) synchronous areas~\cite{Pillai2010}.
Islands serve as a bedrock to study the effects of low-inertial systems, particularly those that rely much more on renewable (non-inertial) energy sources~\cite{RydinGorjao2020a,Milano2018}. 
Furthermore, islanded grids, such as natural islands or islanded microgrids, could play a more important role in the future, e.g. when large synchronous areas are split into smaller areas~\cite{Lidula2011, Parhizi2015}. Such splitting into smaller cells may prevent large-scale cascading failures~\cite{Lagrange2020}.
Within the present article, we will evaluate data recorded in Ireland, Iceland, and the Balearic Islands (see Fig.~\ref{fig:1}).
Ireland is of particular interest here because of its high share of wind energy generation, reaching 43\% of annual generation in 2020~\cite{soni}.
Thereby, it could act as an inspiration for how highly renewable systems can be operated.
Both Ireland and the Balearic Islands have DC connections to larger regions, namely Ireland is connected to Great Britain~\cite{Putz2022}, while the Balearic Islands are connected to mainland Spain (and thereby Continental Europe). 
Meanwhile, Iceland is isolated without any connection to another grid and has a unique generation and demand mixture of geothermal and wind power, as well as data centres and aluminium plants~\cite{iceland}.

\subsection*{Literature review}
The study of islands and (islanded) microgrids has received great interest~\cite{electronics11050791,CHENG2021102856,balearic_HVDC}.
Meanwhile, the area of stochastic modelling of the power-grid frequency has attracted much attention from a broad interdisciplinary audience~\cite{Siegert1998, Friedrich2011, Kwapien2012, Wang2012,Mele2016, Tabar2019, Hossain2019}. 
Fokker--Planck equations have been used to obtain a stochastic description of the observed dynamics~\cite{Schaefer2018b, Vorobev2019}, leading to data-based models~\cite{RydinGorjao2020a} and quantitative comparisons for continental synchronous areas~\cite{Anvari2020}. 
These Fokker--Planck-based approaches have been recently further refined~\cite{Kraljic2022} for short time series.
Complementary, machine learning might assist in estimating suitable parameters~\cite{Kruse2022} or renewable energy generation can be modelled using the same mathematics~\cite{Junlakarn2022}.

\subsection*{Structure}
The present article is structured as follows.
We first introduce the stochastic modelling approach via a Fokker--Planck equation and our newly proposed models in Sec.~\ref{sec:methods}. 
We then demonstrate how our approach reproduces key characteristics of empirical power-grid frequency statistics in Sec.~\ref{sec:results}. 
We continue with a discussion of our results in Sec.~\ref{sec:discussion} and close with an outlook in Sec.~\ref{sec:conclusion}.

\begin{figure}[t]
    \includegraphics[width=\columnwidth]{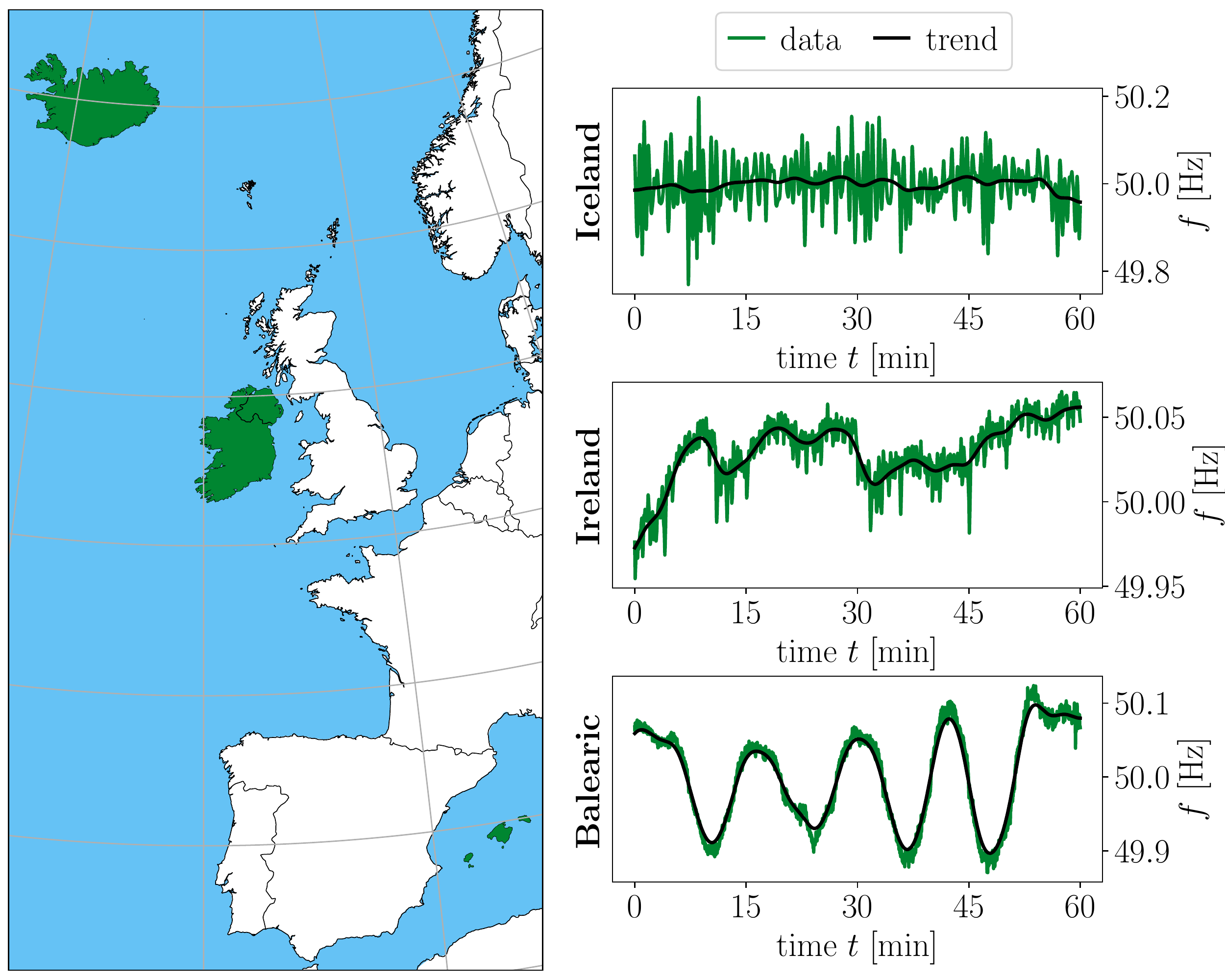}
    \caption{Power-grid frequency recordings from Iceland, Ireland, and the Balearic Islands.
    Black lines indicate the Gaussian filter used for detrending.}\vspace{-0.3cm}\label{fig:1}
\end{figure}

\section{Methods} \label{sec:methods}
\subsection*{Deriving models}
In this work, we approximate the dynamics of the power-grid frequency via stochastic differential equations (SDEs)~\cite{Gardiner2009, Risken1996}, based on the equation-of-motion of the aggregated swing equation~\cite{Schaefer2018b, RydinGorjao2020a,Ulbig2014, Schmietendorf2017} for the bulk angular velocity $\omega$ and bulk angle $\theta$ of a power network, given by
\begin{equation}
\begin{aligned}\label{eq:swing}
    \frac{\text{d}\theta}{\text{d}t} &= \dot \theta =\omega,\\
    \frac{\text{d}\omega}{\text{d}t} &= \dot \omega = c_1(\omega) + c_2(\theta,\omega) + \Delta P + \epsilon(\omega)\xi,
\end{aligned}
\end{equation}
with $c_1(\omega)$ the primary or droop control function, $c_2(\theta,\omega)$ the secondary or integral control function, $\Delta P$ the power mismatch, $\epsilon(\omega)$ the noise amplitude (potentially dependent on $\omega$), and $\xi$ Gaussian white noise.
We note that this stochastic swing equation has been normalised by an unknown inertial constant $M$.
The (bulk) angular velocity is connected to the power-grid frequency as $f=f_0+\frac{\omega}{2 \pi}$, with reference frequency $f_0= 50\,$Hz in our case.
Hence, frequency deviations ($f-f_0$) are proportional to the angular velocity $\omega$.
The power mismatch $\Delta P$ represents all deterministic influences on the grid's power balance, in particular, due to power dispatch: At regular intervals, typically every $15$ to $60$ minutes, the scheduled generation is updated to the continuously changing load.
These discrete updates induce deterministic frequency deviations~\cite{Schaefer2018c,Kruse2021b}, which we model via $\Delta P$.
The primary-control function $c_1(\omega)$ and the integral secondary-control function play a role in restoring power-grid angular velocity to its nominal value~\cite{Machowski2020}.
The white noise process $\xi$ may be interpreted as the time derivative of a Wiener process $\text{d}W/\text{d}t=\xi$ and accounts for all high-frequency fluctuations and microscopic noise observed in a power grid~\cite{RydinGorjao2022}.

As this is a stochastic system, we have to rely on probabilistic results, i.e., instead of a trajectory of $\omega$, we report and model the evolution of the probability density function $\rho(\theta,\omega)$ of the rotor-angle $\theta$ and the angular velocity $\omega$ via a Fokker--Planck equation based on~\cite{RydinGorjao2020a, Acebron1998, Acebron2000}
\begin{equation}\label{eq:fokker-planck2D}
    \frac{\partial \rho}{\partial t} = 
    - \frac{\partial }{\partial \theta} (\omega\rho)
    -\frac{\partial }{\partial \omega}\left((c_1(\omega) + c_2(\theta, \omega) \right)\rho)  +  \frac{\partial^2 }{\partial \omega^2}\left (\frac{\epsilon(\omega)^2}{2} \rho \right).
\end{equation}
Assuming that secondary control acts on a different time scale, we might neglect the $\theta$ dynamics in the estimation of $c_1(\omega)$ and $\epsilon(\omega)$ and simply consider the 1D case:
\begin{equation}\label{eq:fokker-planck1D}
    \frac{\partial \rho}{\partial t} = -\frac{\partial }{\partial \omega}\left(c_1(\omega)\rho\right) +  \frac{\partial^2 }{\partial \omega^2}\left (\frac{\epsilon(\omega)^2}{2} \rho \right).
\end{equation}
Previous works have focused solely on solving the 1D Fokker--Planck equation~\eqref{eq:fokker-planck1D} by neglecting the $\theta$ dynamics and effectively obtaining an expression for $\rho(\omega)$~\cite{RydinGorjao2020a, Kraljic2022, RydinGorjao2022}.
As a further simplification, previous work assumed that the primary control is fully linear $c_1(\omega)\sim c_1\omega$ and that noise is purely additive $\epsilon(\omega)\sim\epsilon$.
Secondary control $c_2$ was then either completely neglected or calculated in a second step.
These simplified models result in an augmented Ornstein--Uhlenbeck SDE, to which we know the explicit closed-form solution~\cite{RydinGorjao2020a}.

\begin{figure}[t]
    \includegraphics[width=\columnwidth]{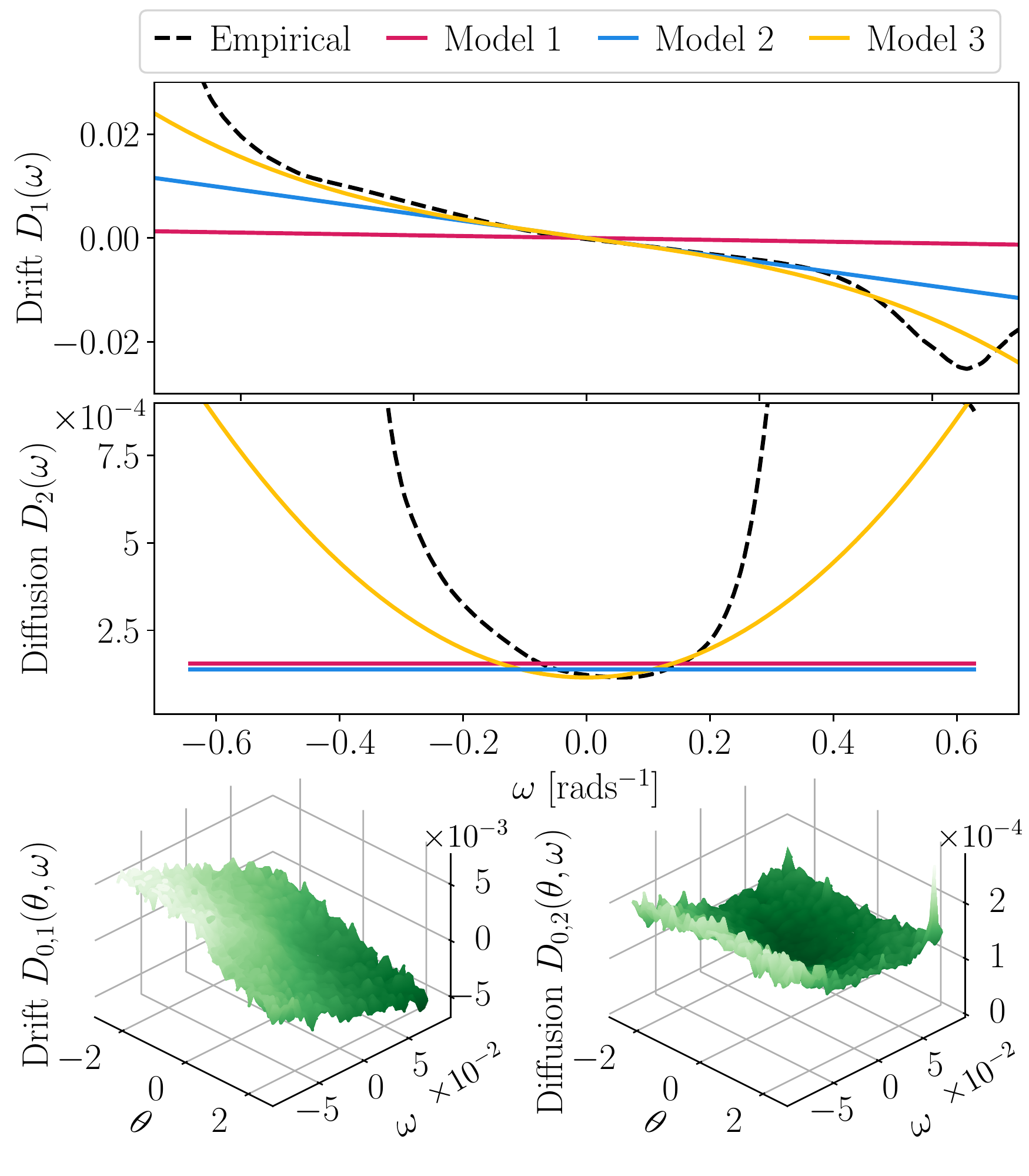}
    \caption{Drift and diffusion from Ireland. 
    Polynomial fits are used in Model~3 for both drift and state-dependent diffusion (top rows).
    For Model 4, we estimate 2D Kramers--Moyal coefficients (bottom row).}\vspace{-0.3cm}\label{fig:2}
\end{figure}

To estimate the control parameters $c_1$, $c_2$, and the noise amplitude $\epsilon$ purely from data, we turn to the Nadaraya--Watson non-parametric kernel-density Kramers--Moyal coefficients $D_n(x)$ estimation~\cite{Nadaraya1964, Watson1964, Lamouroux2009, RydinGorjao2019}, which reads:
\begin{equation}
\begin{aligned}
    D_{\vec{n}}(\vec{x}) &\sim \frac{1}{n!}\frac{1}{\Delta t}\langle (\vec{x}(t+\Delta t) - \vec{x}(t))^n|\vec{x}(t) = \vec{x} \rangle \\
    & \sim \frac{1}{n!}\frac{1}{\Delta t}\frac{1}{N}\sum_{i=1}^{N-1}(\vec{x}_{i+1}-\vec{x}_{i})^{\vec{n}} K(\vec{x} - \vec{x}_i),
\end{aligned}
\end{equation}
where $\vec{x}$ is either $\omega$ for the 1D case or $(\omega, \theta)$ for the 2D case. $N$ is the number of data points from a timeseries, $K$ is a kernel with bandwidth $h$ (c.f.~\cite{Lamouroux2009, RydinGorjao2019}), and $\Delta t$ is the sampling rate.
The order of the Kramers--Moyal coefficient $\vec{n}$ depends on the dimension of $\vec{x}$: in 1D, $\vec{n}=n$ is an integer; in 2D, $\vec{n}=(n,m)$ is a tuple of two integers. 
Therein we estimate $c_1$, $c_2$, and $\epsilon$ directly from data, in either a 1D or a 2D setting (see Fig.~\ref{fig:2}).
The Kramers--Moyal coefficients allow us to disentangle the Fokker--Planck equations as given in~\eqref{eq:fokker-planck2D} and~\eqref{eq:fokker-planck1D}.
We have, in 1D~\cite{Tabar2019,Gardiner2009,Risken1996}:
\begin{equation}
\begin{aligned}
    D_1(\omega) &=c_1(\omega),\\
    D_2(\omega) & =\epsilon(\omega)^2/2,
\end{aligned}
\end{equation}
and in 2D:
\begin{equation}
\begin{aligned} \label{KM_Coeff_2d}
    D_{1,0}(\theta,\omega) &\approx \omega,\\ 
    D_{0,1}(\theta,\omega) &= c_1(\omega)+c_2(\theta),\\
    D_{0,2}(\theta,\omega) & = \epsilon(\omega)^2/2.
\end{aligned}
\end{equation}
The estimated coefficients $D_1$, $D_2$, as well as $D_{0,1}$ and $D_{0,2}$ can be intricate functions of $\theta$ and $\omega$, see Fig.~\ref{fig:2}.
We note that \textit{a priori} we could consider a noise term in the rotor-angle $\theta$ as well, which would lead us to investigate $D_{2,0}$.
However, due to the underlying equation of motion $\dot \theta =\omega$, any noise in $\theta$ would in turn result in a measurable noise in $\omega$, which we observe via measurements in the frequency $f$. Therefore, we consider only noise in $\omega$.

Data for all three islands (Iceland, Ireland, Balearic islands) are recorded via the electrical data recorder (EDR), see \cite{maass2013first, RydinGorjao2020b, Jumar2020} for details, and frequency measurements $f$ with a time resolution of $1$ second or $0.1$ seconds are available for several weeks.
We separate the power-grid dynamics into a slowly moving trend (described by $\Delta P$ and $c_2$, as explained in \cite{RydinGorjao2020a}) and short-term fluctuations (captured mostly by $c_1$ and $\epsilon$) by applying a Gaussian filter with a window of $60$~seconds 
(see Fig.~\ref{fig:1}).

\begin{figure*}[t]
    \centering
    \includegraphics[width=0.9\textwidth]{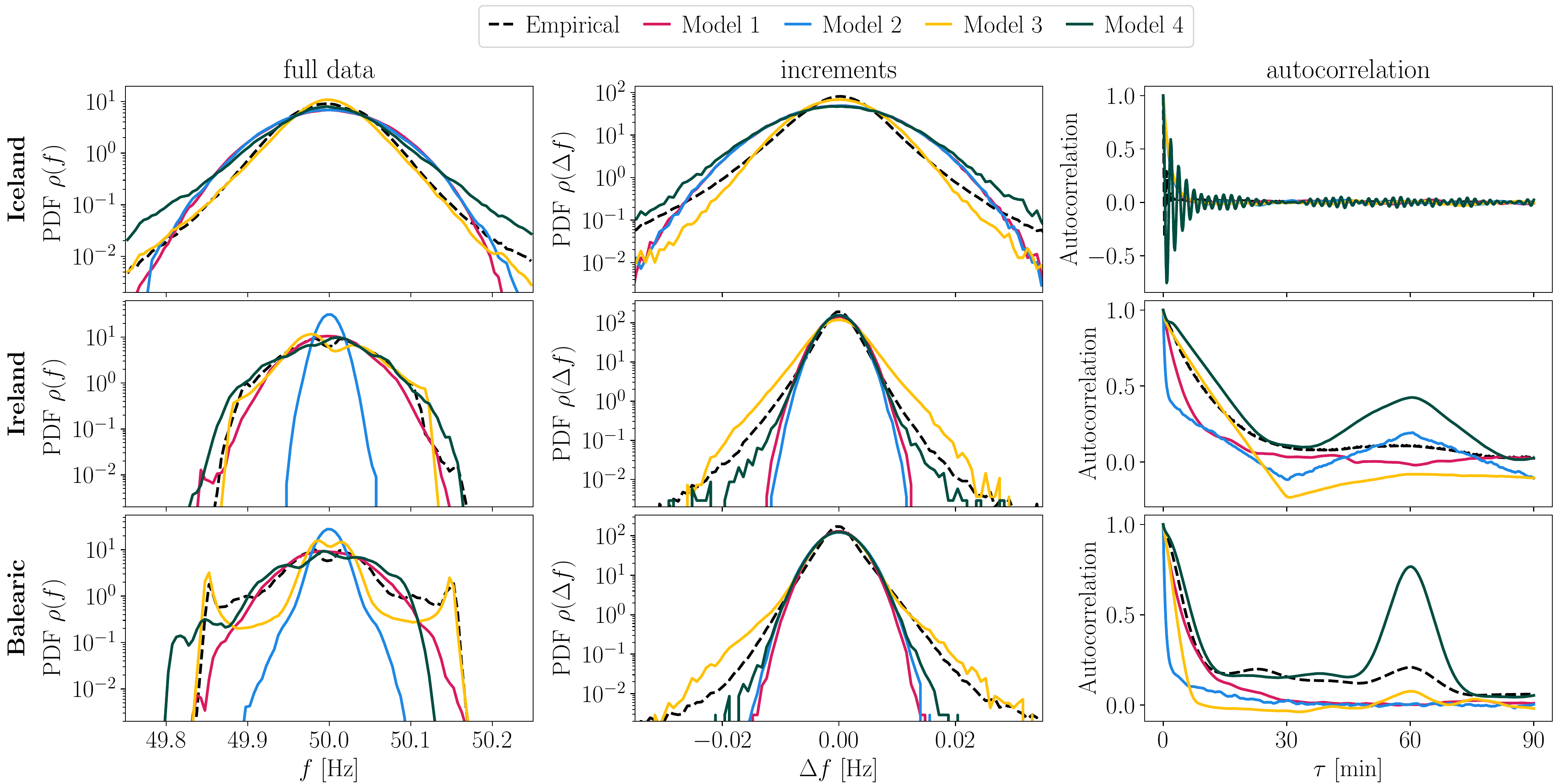}
    \caption{Probability density of the frequency $f$ (left), its increments $\Delta f$ (center), and the autocorrelation function for $90$ minutes (right).}\vspace{-0.3cm}\label{fig:3}
\end{figure*}

\subsection*{Models}
In this article, we compare four models, two reference cases and two expansions of previous models:
\begin{itemize}
    \item \textbf{Model 1 (reference)}: Basic  Ornstein--Uhlenbeck process, only one damping constant and noise:
    \begin{equation}\label{eq:model1}
        \dot \omega = c_1\omega + \epsilon\xi,
    \end{equation}
    where the constant values for $c_1$ and $\epsilon$ are estimated directly from the original data using the Kramers--Moyal coefficients, without any detrending. Note that here the $\epsilon$ has to include all deterministic and stochastic variations of the frequency, while $c_1$ has to capture all restoring forces (primary control, secondary control, deterministic relaxation).
    
    \item \textbf{Model 2 (reference)}: The linear-response model employed in~\cite{RydinGorjao2020a}:
    \begin{equation}\label{eq:model2}
        \dot \theta =\omega, \quad \dot \omega = c_1\omega + c_2\theta + \Delta P + \epsilon\xi.
    \end{equation}
    The constants $c_1$ and $\epsilon$ are derived by detrending the time series and estimating the drift and diffusion using the Kramers--Moyal coefficients~\cite{RydinGorjao2019}.
    The secondary control constant is then estimated by combining the trajectory and the estimate of $c_1$, while the deterministic dynamic is included via a time-dependent Heaviside function as the power mismatch $\Delta P$, see~\cite{RydinGorjao2020a} for details.
    
    \item \textbf{Model 3 (our contribution)}: Our first extended model with non-linear response and multiplicative noise~\cite{RydinGorjao2022}:
    \begin{equation}\label{eq:model3}
        \dot \theta =\omega, \quad \dot \omega = c_1(\omega) + c_2(\theta,\omega) + \Delta P + \epsilon(\omega)\xi,
    \end{equation}
    wherein we non-parametrically approximate $c_1(\omega)$ and $\epsilon(\omega)$ from the detrended time series. 
    Fitting the empirical functions more closely, we use a polynomial of order $3$ for the drift and a quadratic function in $\omega$ to describe the diffusion, see Fig.~\ref{fig:2}.
    The power mismatch $\Delta P$ is modelled by a time-dependent Heaviside function.
    These steps, mimicking power dispatch and trading activities, take place every $30$ minutes for the Irish grid and every $60$ minutes for the Balearic grid, while no such steps are included for the flat Icelandic profile. 
    We approximate the power step as the change in frequency $\Delta P \approx \dot \omega$ in an appropriate time interval around changes of power feed-in. 
    As in Model 2, we estimate the secondary control $c_2(\theta,\omega)$ by solving equation \eqref{eq:model3} neglecting the noise (setting $\epsilon =0$), where we approximate the non-linear primary control $c_1(\omega) = q_3\omega^3 + q_1\omega$ by its first-order Taylor polynomial $c_1^\text{Taylor}(\omega)$.
    Further, after a change of the deterministic power dispatch, the frequency jumps and then decays back approximately exponentially following $\exp(-t/\tau)$. 
    Plugging in the Taylor expansion for $c_1(\omega)$, this is further simplified to $1/\tau\approx c_2(\theta,\omega)/\left(c_1^\text{Taylor}(\omega) \theta \right)$~\cite{RydinGorjao2020a}.
    Hence, we use a first-order expansion of $c_1(\omega)$ to compute $c_2(\theta, \omega)$:
    \begin{align}
        c_2(\theta,\omega) \approx \frac{1}{\tau } \left(3q_3\omega^2 + q_1 \right)\theta.
    \end{align}
    In addition, for the Balearic and the Irish grids, we increase the primary control by a factor $3$ for high-frequency deviations ($|f-f_0| 	\gtrsim 150\,$mHz), to mimic power exchange via HVDC lines~\cite{Putz2022,balearic_HVDC}.

    \item \textbf{Model 4 (our contribution)}: Our second extended model separates the frequency into stochastic fluctuations and a trend:
    \begin{align}\nonumber
        \dot \theta_\text{fluct} &=\omega_\text{fluct}, \\  \nonumber 
        \dot \omega_\text{fluct} &= c_1(\omega_\text{fluct}) + c_2(\theta_\text{fluct}) + \epsilon(\omega_\text{fluct})\xi \\
        \omega &= \omega_\text{fluct} + \omega_\text{trend},
    \end{align}\label{eq:model4}where $\omega_\text{trend}$ is given by a multiple of a Gaussian-filtered daily profile. 
    In this model, we focus on the stochastic dynamics of the frequency and consider a strengthened filtered daily profile as the deterministic drive of the model in order to recreate the entire dynamics. The multiplication is necessary to receive a suitable width of the distribution as the daily profile averages over the large deterministic fluctuations. \\
    In contrast to Models 2 and 3, we estimate the secondary control $c_2(\theta)$ directly by calculating the Kramers--Moyal coefficients from the bivariate (2D) Fokker--Planck equation \eqref{eq:fokker-planck2D}.
    In order to obtain a time series for the voltage angle $\theta$, we integrate over the detrended angular velocity $\omega$.
    The control $c_1$, $c_2$, and noise $\epsilon$ can therefore be estimated from the Kramers--Moyal coefficients ~\eqref{KM_Coeff_2d}. For simplicity, we assume $c_1(\omega)\sim \omega$, $c_2(\theta)\sim \theta$, $\epsilon(\omega)\sim \sqrt{\text{const.}+\omega^2}$.
    Furthermore, as in Model 3, an increase of $c_1$ for high-frequency deviations simulates the influence of the HVDC response.
\end{itemize}

We note that the parameters $c_1$, $c_2$, and $\epsilon$ are not identical between models but carry a similar function, i.e. they symbolise primary control, secondary control, and fluctuation amplitude, respectively.
For the implementation details see~\cite{github}.

\subsection*{Quantifying models}
The target of the present article is to adequately reproduce the statistics of power-grid frequency recordings of any generic power grid with minimal information, i.e., with an almost purely data-driven approach.
How well does this method perform when applied to islanded grids?

To evaluate the quality of a synthetic-generated probability density function $\rho_{\mathrm{syn}}(x)$ against an empirical one $\rho_{\mathrm{emp}}(x)$, we employ the Kullback--Leibler divergence $D_{\mathrm{KL}}(\rho_{\mathrm{emp}} | \rho_{\mathrm{syn}})$~\cite{Kullback1951ams}

\begin{equation}
    D_{\mathrm{KL}}\left(\rho_{\mathrm{emp}} \mid \rho_{\mathrm{syn}}\right)=\int \rho_{\mathrm{emp}}(x) \ln \left[\frac{\rho_{\mathrm{emp}}(x)}{\rho_{\mathrm{syn}}(x)}\right] \mathrm{d} x.
\end{equation}
A smaller $D_{\mathrm{KL}}$ value implies a better fit between the synthetic and the empirical distributions.

Moreover, given the importance of understanding the response times of the power-grid frequency, we also evaluate the autocorrelation function of the empirical and synthetic data, as given by:
\begin{equation}\label{eq:AC}
    C_{x}(\tau) = \sigma^{-2}\langle(x_{t} - \mu) (x_{t+\tau} - \mu) \rangle,
\end{equation}
where $x$ is a time series (frequency or frequency increments, empirical or synthetic), $\mu$ is the mean value of $x$, and $\sigma$ is the standard deviation of $x$.

Code to reproduce the results is available online~\cite{github} and data are freely available, see \cite{RydinGorjao2020b, Jumar2020} and \url{www.power-grid-frequency.org}.

\section{Results} \label{sec:results}

Let us review the results in three steps: What can we learn about the dynamic and statistical properties of the islands?
Which characteristics are reproduced by the models?
How do the different models perform quantitatively?

\subsection*{Characteristics of the data}
The frequency statistics of islands are quite complex and more intricate than in continental regions.
Simply inspecting the empirical data (black lines in Fig.~\ref{fig:3}), we note 
\begin{itemize}
    \item cut-offs for the absolute frequency deviations in Ireland and the Balearic Islands. 
    These likely arise because these islands can balance their power via HVDC lines connected to large synchronous areas. 
    \item highly non-Gaussian statistics, both in the frequency and in the increments.
    For reference, a Gaussian statistic would be indicated via an inverted parabola.
    \item complex autocorrelation functions that decay very rapidly (Iceland) or display more regular peaks (Balearic).
\end{itemize}

\subsection*{Characteristics of the models}
The proposed models capture some of the empirical characteristics, depending on their complexity.

Model 1 by construction only induces Gaussian frequency and Gaussian increment distributions.
Hence, it misses the cut-off for large values and the heavy tails in both frequency and increments.
The autocorrelation decays exponentially but misses the peaks caused by deterministic influences.

Model 2 includes deterministic power mismatch which is not adapted to the characteristics of the grids and therefore is too small for both the Irish and the Balearic grid.
As in Model 1, the increments are Gaussian, missing the heavy tails.
The autocorrelation function decays approximately exponentially with some visibility of deterministic effects, particularly in Ireland. 

Model 3 reproduces the multi-modal distributions in Ireland and the Balearic islands and even includes frequency cut-offs at large values.
These characteristics are possible due to the cubic primary control and an even stronger control for large deviations, e.g. at $|f|\gtrsim 150\,$mHz in the Balearic grid.
The increments display heavy tails, as in the real data, due to multiplicative noise, i.e. $\epsilon(\omega)$ being explicitly state-dependent in this model.
The autocorrelation function reproduces the decay and some small peaks at the $60$-minute mark.

Model 4 has some weaknesses in approximating the empirical probability density and the autocorrelation function, as the latter exhibits large peaks at the $60$-minutes mark.
This is mostly due to the heuristic estimation of the frequency trend. 
Meanwhile, the increments follow the main characteristics of the empirical data. 
As in Model 3, the multiplicative noise $\epsilon(\omega)$ (state-dependent) facilitates non-Gaussian increments.

\subsection*{Comparison of performance}
\begin{figure}[t]

  \includegraphics[width=\columnwidth]{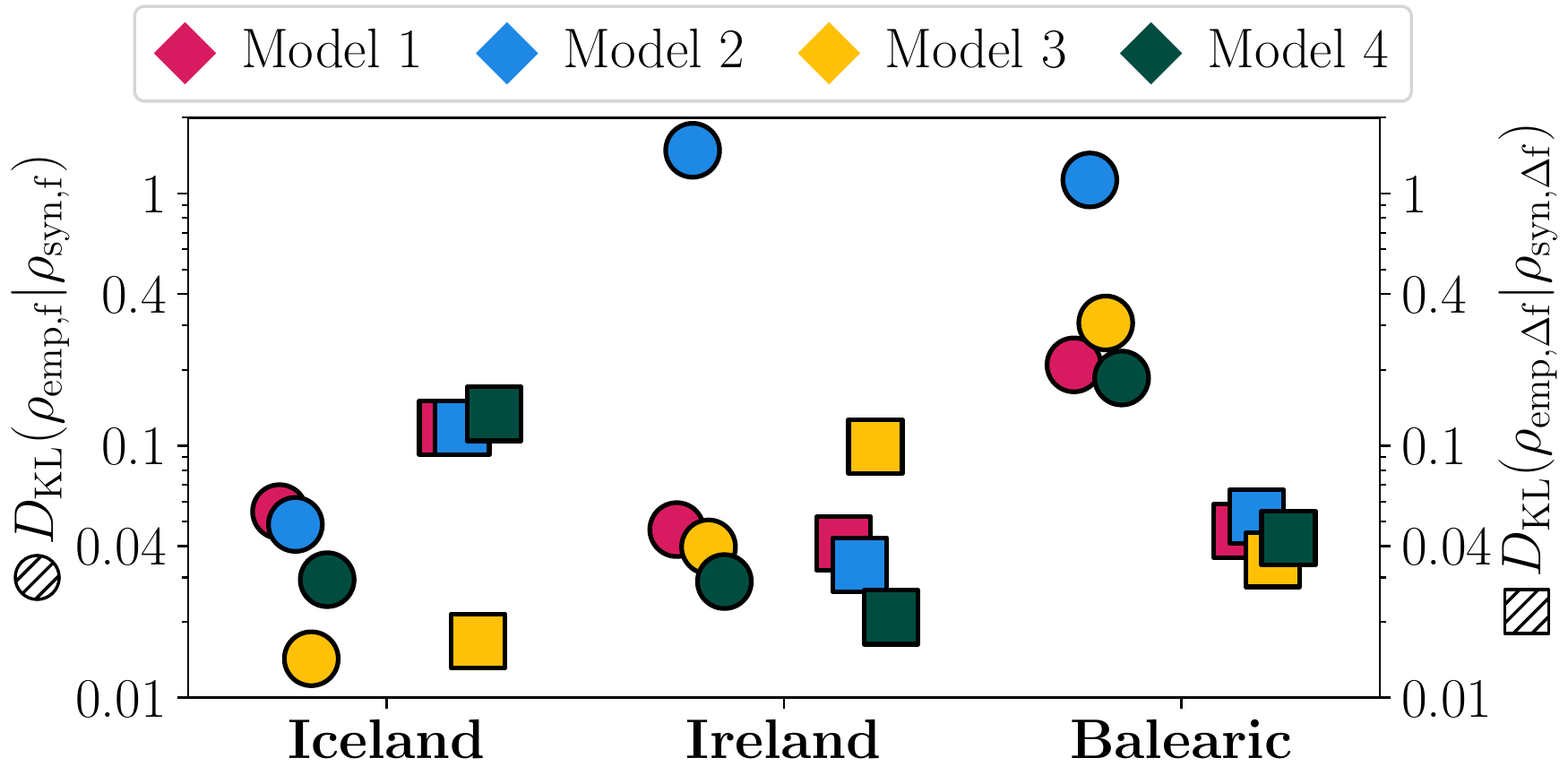}
  \caption{Kullback--Leibler divergence between the empirical and 4 generated synthetic PDFs for the frequency (left) and the increments (right).}\vspace{-0.3cm}\label{fig:4}
\end{figure}

Going beyond the qualitative comparison of the characteristics, we also compare how well the different models quantitatively fit the empirical data, measured via the Kullback--Leibler divergence, see Fig.~\ref{fig:4}.
We note that the standard Ornstein--Uhlenbeck process (Model 1) always provides a decent description of the frequency statistic (circles) and, by design, matches the empirical standard deviation well.
Meanwhile, it tends to be among the worst performers in the increment analysis (squares), as it can only reproduce Gaussian increments. This oversimplification becomes most apparent when investigating the increment tails (Fig.~\ref{fig:3}).
The previously developed data-driven model~\cite{RydinGorjao2020a} (Model 2) encounters problems when applied to islands without any adjustments.
In particular, the overall probability distribution (circles) is among the poorest-performing models for all three islands. 

The analysis of the two new models both shows that these are promising developments but that there still remains potential for improvement. 
Model 3 performs very well in the Icelandic data, while Model 4 is the best model for Ireland.
In particular in the increment analysis (squares), one of the new models is always the best-performing one.
This indicates two points: First, our advanced modelling is especially useful for modelling the stochastic dynamics, as seen in their increments statistics. Second, a fully generalised model, applicable to data from continents (as previously done with Model 2) and diverse islands (as done here) is not yet available.

\section{Discussion} \label{sec:discussion}

Overall, we show that stochastic power-grid frequency models, aided by a Fokker--Planck description of the underlying physical process, can reproduce the statistics, increments, and two-point correlation of power-grid frequency recordings from various grids without access to in-depth information of each power-grid system's network details.
We show that solely from power-grid frequency recordings, we can estimate the strength of primary and secondary control as well as the amplitude of the noise or high-frequency fluctuations.
Armed only with these data-driven functions, we can construct stochastic differential equations that reproduce both the statistics as well as the autocorrelation structure of a large class of power grids.
In this work, in particular, we go one step further and examine the Icelandic, Irish~\cite{Mele2016}, and Balearic power grids, which have gathered far less attention in the scientific community than other major grids, such as Continental Europe~\cite{RydinGorjao2020a, RydinGorjao2020b} or Great Britain~\cite{Vorobev2019, Kraljic2022}.
The two new models offered in this work -- a one-dimensional, non-linear and a two-dimensional Fokker--Planck model -- consistently approximate the increments statistics of the frequency very well. 
Meanwhile, the autocorrelation and aggregated frequency statistics are much harder to describe. 

\section{Conclusion and Outlook} \label{sec:conclusion}
Understanding the dynamics and statistics of these size-wise smaller power grids is crucial for the understanding of insular grids, which themselves are small-size networks and do not necessarily obey a large-size linear control mechanism observed and understood in major grids.
A representation of power-grid frequency dynamics in a Fokker--Planck setting is central to the diagnosis of irregularities in any power grid.
It offers firstly an understanding of the statistics and therein the effects and cost of control as well as the duration that frequency excursions break out of statutory frequency limits.
Representing power-grid frequency via partial differential equations 
also permits representing frequency as a stochastic differential equation.
This, in turn, allows for the generation of synthetic time series, which can be studied as objects on their own.
Realistic synthetic time series open the door to studying these power grids with data-intensive methods, such as artificial intelligence.

The usage of a cubic polynomial in model 3 recreates the effect of a primary frequency control deadband~\cite{Vorobev2019, Quint} and together with the effect of deterministic power mismatch explains the bimodal distributions in the power-grid frequency statistics.


We present an extension of stochastic modelling, which should be further enhanced in the future. 
Aside from applying our method to more data from different synchronous areas, a more detailed and realistic extraction and modelling of the deterministic power mismatch will be important to better describe empirical data.
We again note that the investigated island grids display a large variation between one another in terms of generation mix and power mismatch.
Hence, it also remains an open challenge to develop a generalised model applicable to various islands or microgrids.

\bibliographystyle{IEEEtran}
\bibliography{bib}

\end{document}